\documentclass[twocolumn]{aastex6}
\usepackage{gensymb}
\usepackage{natbib}
\begin{document}
\turnoffedit{}
\submitted{}
\title{H$_2$O$_2$ within chaos terrain on Europa's leading hemisphere}

\author{Samantha K. Trumbo, Michael E. Brown}
\affil{Division of Geological and Planetary Sciences, California Institute of Technology, Pasadena, CA 91125, USA}
\author{Kevin P. Hand}
\affil{Jet Propulsion Laboratory, California Institute of Technology, Pasadena, CA 91109, USA}

\begin{abstract}

Hydrogen peroxide is part of an important radiolytic cycle on Europa and may be a critical source of oxidants to the putative subsurface ocean. The surface geographic distribution of hydrogen peroxide may constrain the processes governing its abundance as well as its potential relevance to the subsurface chemistry. However, maps of Europa's hydrogen peroxide beyond hemispherical averages have never been published. Here, we present spatially resolved L-band (3.16--4 $\micron$) observations of Europa's 3.5 $\micron$ hydrogen peroxide absorption, which we obtained using the near-infrared spectrometer NIRSPEC and the adaptive optics system on the Keck II telescope. Using these data, we map the strength of the 3.5 $\micron$ absorption across the surface at a nominal spatial resolution of $\sim$300 km. Though previous disk-integrated data seemed consistent with the laboratory expectation that Europa's hydrogen peroxide exists primarily in its coldest and iciest regions, we find nearly the exact opposite at this finer spatial scale. Instead, we observe the largest hydrogen peroxide absorptions at low latitudes on the leading and anti-Jovian hemispheres, correlated with chaos terrain, and relative depletions toward the cold, icy high latitudes. This distribution may reflect the effects of decreased hydrogen peroxide destruction due to efficient electron scavenging by CO$_2$ within chaos terrain.
\end{abstract}


\section{Introduction}\label{sec:intro}
The continuous bombardment of Europa's surface by energetic particles within Jupiter's magnetosphere results in extensive radiolytic processing of the surface ice \citep[e.g.][]{JohnsonQuickenden1997}. The impinging electrons, protons, and ions lead to the dissociation of water molecules, producing OH radicals, which can recombine to produce hydrogen peroxide (H$_2$O$_2$) and other oxidants \citep[e.g][]{JohnsonEtAl2003, CooperEtAl2003, LoefflerEtAl2006}. H$_2$ is lost in the process, creating an increasingly oxidizing surface. Understanding this radiolytic cycle on Europa is not only important in terms of Europa's surface composition and the general study of surface-magnetosphere interactions on icy bodies, but it is also critical for assessing the potential chemical energy sources within Europa's subsurface ocean. Hydrothermal processes at the seafloor may be a source of reductants, but the potential habitability of Europa's ocean will likely depend on a complementary supply of oxidants, such as H$_2$O$_2$, from the radiolytically processed surface \citep{Chyba2000,Hand2009}.

The geographic distribution of Europa's H$_2$O$_2$ may hold clues to the processes controlling its presence on the surface as well as to its potential delivery to the subsurface ocean. The \textit{Galileo} Near Infrared Mapping Spectrometer (NIMS) definitively detected H$_2$O$_2$ on the surface via a prominent 3.5 $\micron$ absorption in an average spectrum of the leading/anti-Jovian quadrant \citep{Carlson1999Perox}. However, the intense radiation encountered during closer flybys of Europa resulted in poor-quality spectra beyond 3 $\micron$, and no spatially resolved NIMS maps of H$_2$O$_2$ were ever published. To date, ground-based spectra, in which the 3.5 $\micron$ absorption of H$_2$O$_2$ was rotationally resolved across four nights of observation, have provided the best constraints on its geography \citep{HandBrown2013}. These spectra show the largest absorptions on the leading and anti-Jovian hemispheres, in agreement with the NIMS observation, and almost no detectable absorption on the trailing hemisphere. At this scale, Europa's H$_2$O$_2$ distribution was easily attributed to the availability of surface water ice and the low surface temperatures associated with Europa's iciest regions, which lie at the mid to high latitudes of the leading and anti-Jovian hemispheres \citep{BrownHand2013}. Water molecules are a necessary precursor for H$_2$O$_2$ production, and laboratory experiments studying the production of H$_2$O$_2$ via water ice radiolysis \citep{MooreHudson2000,LoefflerEtAl2006, ZhengEtAl2006, HandCarlson2011} consistently observe lower equilibrium abundances with increasing temperature, in some cases by nearly a factor of seven across the $\sim$80--120 K temperature range relevant to Europa \citep{HandCarlson2011}.

Considering all of the published observations and laboratory data, one might reasonably expect that higher spatial resolution observations show an abundance of H$_2$O$_2$ at the icy high latitudes of Europa's leading and anti-Jovian hemispheres, and relative depletions at the warm equator, which likely contains a significant non-ice fraction in the form of \edit1{endogenous} salt deposits associated with \edit1{largescale regions of geologically young and highly disrupted} chaos terrain \citep{Fischer2015, Fischer2017, TrumboEtAl2019}. In order to address this hypothesis, we obtained spatially resolved L-band spectra of Europa's surface, with the goal of mapping the geographic distribution of its 3.5 $\micron$ H$_2$O$_2$ band. 

\section{Observations and Data Reduction}\label{sec:observations}
We observed Europa on  Feb. 24th and 25th of 2016 and again on June 6th of 2018 using the near-infrared spectrograph NIRSPEC and the adaptive optics system on the KECK II telescope. We used the 3.92$^{\prime \prime}$ x 0.072$^{\prime \prime}$  slit in low-resolution mode (R $\sim$ 2000). Our data cover the wavelength range of $\sim$3.16--4 $\micron$ in the L-band. As a telluric calibrator for the 2016 observations, we observed HD 98947, a V = 6.9 G5 star that was $\sim$ 1.2$\degree$ from Europa on the sky. In 2018, we used HD 128596, a V = 7.5 G2 star with a 3.9$\degree$ separation from Europa. We observed all targets in an ABBA nodding pattern. For our 2016 data, each Europa pointing consisted of 30 20-second coadds, and each calibrator pointing consisted of  15 2-second coadds. In 2018, we reduced the integration times to 30 10-second co-adds for each Europa pointing and 10 2-second coadds for each calibrator pointing. 

\floattable
\begin{deluxetable}{cccccccc}
\tablecaption{Table of Observations\label{table:obs}}
\tablecolumns{6}
\tablenum{1}
\tablewidth{0pt}
\tablehead{
\colhead{Date} & \colhead{Hemisphere} & \colhead{Slit} & \colhead{Airmass} & \colhead{Total Int.} & \colhead{Central} &\colhead{Central} & \colhead{Telluric} \\
\colhead{(UT)} & \colhead{ } & \colhead{Orientation} & \colhead{Range} & \colhead{Time (min)} &  \colhead{Longitude} & \colhead{Latitude} & \colhead{Calibrator}}

\startdata
2016 Feb 24 & trailing/sub-Jovian & E/W & 1.10--1.75 & 120 & 338 W & 1 S & HD 98947\\
2016 Feb 24 & sub-Jovian & N/S & 1.03--1.07 & 100 & 348 W & 2 S & HD 98947\\
2016 Feb 25 & leading & E/W & 1.24--1.38 & 40 & 79 W & 0 N & HD 98947\\
2016 Feb 25 & leading & E/W & 1.11--1.18 & 40 & 82 W & 12 S & HD 98947\\
2016 Feb 25 & leading & N/S & 1.04--1.08 & 40 & 88 W & 2 S & HD 98947\\
2016 Feb 25 & leading & E/W & 1.08--1.14 & 40 & 95 W & 0 N & HD 98947\\
2018 Jun 06 & leading & N/S & 1.40--1.42 & 10 & 119 W & 4 S & HD 128596\\
2018 Jun 06 & leading & N/S & 1.36--1.38 & 10 & 124 W & 4 S & HD 128596\\
2018 Jun 06 & leading/anti-Jovian & N/S & 1.24--1.30 & 20 & 166 W & 3 S & HD 128596\\
2018 Jun 06 & leading/anti-Jovian & N/S & 1.22--1.26 & 20 & 146 W & 3 S & HD 128596\\
2018 Jun 06 & leading/anti-Jovian & N/S & 1.25--1.25 & 10 & 162 W & 3 S & HD 128596\\
2018 Jun 06 & leading & N/S & 1.22--1.22 & 15 & 115 W & 4 S & HD 128596\\
2018 Jun 06 & leading & N/S & 1.23--1.25 & 20 & 104 W & 3 S & HD 128596\\
2018 Jun 06 & leading & N/S & 1.26--1.30 & 20 & 87 W & 3 S & HD 128596\\
2018 Jun 06 & leading & N/S & 1.30--1.32 & 10 & 120 W & 4 S & HD 128596\\
\enddata
\end{deluxetable}

During both sets of observations, Europa had an angular diameter of nearly 1$^{\prime \prime}$, corresponding to $\sim$ 10 ~300-km resolution elements at the diffraction limit of Keck at 3.5 $\micron$. For each Europa pointing, we aligned the slit in either an E/W or N/S orientation with respect to Europa's north pole. We obtained a corresponding SCAM guide camera image for each exposure as a record of the exact slit positions on the disk. The telescope pointing tended to drift significantly during nods for the 2018 observations. Thus, we used these images in real time to manually maintain a consistent slit position across each ABBA set. 

We reduce all data in Python, following the standard methodology of image rectification, image pair subtraction, residual sky subtraction, and telluric and wavelength calibration, utilizing both the Astropy \citep{Astropy} and skimage.transform \citep{skimage} packages. We use an ATRAN atmospheric transmission spectrum \citep{ATRAN} for wavelength calibration. Some of the raw spectral files display readout artifacts, which artificially brighten every 8th pixel in the dispersion direction. Thus, we replace each affected pixel by the average of the two immediately adjacent pixels.

Following this initial data reduction, we extract spectra for individual spatial resolution elements within each slit position. Using the Python Basemap package and the calculated size of Europa in SCAM pixels (0.0168$^{\prime \prime}$/pixel), we align the SCAM image corresponding to each NIRSPEC exposure with an orthographic projection and estimate the coordinates of the slit on the disk. Though we made real-time manual corrections to the inaccurate telescope nodding experienced during the 2018 observations, some ABBA sets still exhibited significant position discrepancies between nods. Thus, we align SCAM images and estimate the corresponding slit coordinates for each nod independently. We then locate Europa in the 2D spectral images using its calculated size in NIRSPEC pixels (0.013$^{\prime \prime}$/pixel) and extract individual spectra corresponding to an 8-pixel spatial resolution element, stepping by 4 spatial pixels between extractions. We determine the geographic coordinates of each extracted spectrum using the previously obtained slit positions and the aforementioned NIRSPEC pixel scale. 

To minimize geographic smearing, while still maximizing the signal-to-noise, we average spectra from overlapping slit positions, as determined by manually inspecting the SCAM images and the extracted coordinates corresponding to each nod. In the case of the trailing/sub-Jovian slits, we tolerate moderate spatial smearing in favor of the enhanced signal-to-noise needed to quantify weak H$_2$O$_2$ absorptions. In all cases, we account for any resulting adjustments to the effective slit width and to the geographic area covered by each spectrum. Table \ref{table:obs} summarizes our observations and the geometries of geographically unique slit positions.

\section{H$_2$O$_2$ Mapping}\label{sec:mapping}
In order to determine the geographic distribution of Europa's 3.5 $\micron$ H$_2$O$_2$ absorption, we calculate its band \edit1{area} in each individual extracted spectrum. For most spectra, we fit a second-order polynomial from 3.37 to 3.715 $\micron$, excluding the portion corresponding to the H$_2$O$_2$ absorption ($\sim$ 3.4--3.65 $\micron$). For the trailing hemisphere observations, we use a third-order polynomial, which better matches the continuum shape. We inspect all fits by eye and, if necessary to achieve a satisfactory continuum fit, make small adjustments to these parameters. We then remove the calculated continuum from each spectrum and integrate the residual absorption to calculate the equivalent width (i.e. the width of a 100\% absorption of the same integrated area). We map the resulting band \edit1{areas} using the slit positions and widths obtained as in Section \ref{sec:observations}. Our 2016 data show generally stronger absorptions than do our 2018 data\edit1{, with maximum band areas $\sim$25\% larger than those observed in 2018}. This is perhaps unsurprising given that H$_2$O$_2$ concentrations on Europa reflect a dynamic equilibrium between constant formation and decay \citep{HandCarlson2011} \edit1{that may be influenced by the temporal variability of the radiation environment or of the local surface temperature}. Thus, as we are concerned primarily with geographic correlations, we map the 2016 and 2018 slits separately.

\begin{figure*}
\figurenum{1}
\plotone{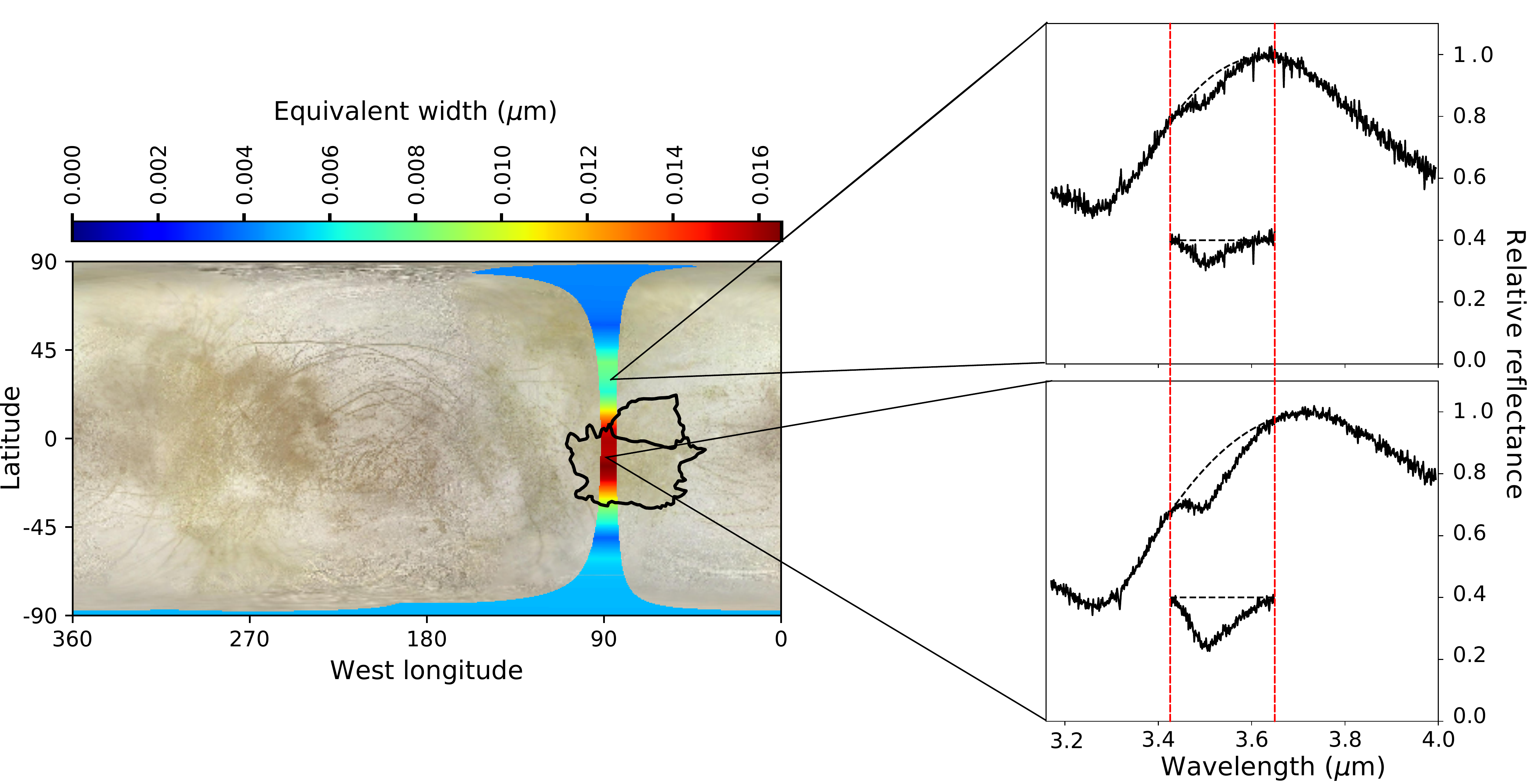}
\caption{A single N/S slit from our 2016 observations, which falls on the leading hemisphere and crosses both the most spectrally icy location on the surface \citep[$\sim$ 30$\degree$ N, 90$\degree$ W,][]{BrownHand2013} and the salty, low-latitude chaos region Tara Regio. Contrary to the hypothesis that Europa's H$_2$O$_2$ should follow the cold, icy terrain of the upper latitudes, the 3.5 $\micron$ band appears anti-correlated with both temperature and ice availability. Instead, the strongest absorptions fall nearly perfectly within the outlined bounds of Tara Regio, with band \edit1{areas} more than twice that of the aforementioned ice-rich region. We show representative spectra from both locations to the right of the map, where the dashed red lines outline the H$_2$O$_2$ band. Second-order polynomial continua are indicated by the dashed black curves. We also include the continuum-removed absorptions to ease comparison of the band strengths, although the differences are readily apparent in the spectra themselves. Both spectra are normalized to their individual peaks in the 3.6--3.7 $\micron$ region. \label{fig:2016_spectra}}
\end{figure*}

Figure \ref{fig:2016_spectra} shows a single mapped slit from our 2016 observations. This slit position crosses both the most spectrally icy location on the surface at $\sim$30$\degree$ N and 90$\degree$ W on the leading hemisphere \citep{BrownHand2013} and the comparatively warm and ice-poor chaos region Tara Regio, making it ideal for testing the hypothesis that Europa's H$_2$O$_2$ prefers cold, icy terrain. Puzzlingly, contrary to this hypothesis, the 3.5 $\micron$ H$_2$O$_2$ absorptions are strongest in exactly the warmest and least icy portion of the slit, nearly perfectly corresponding to the bounds of Tara Regio (mapped approximately from \citet{Doggett2009}). \edit1{Tara Regio is located at low latitudes, where daytime temperatures approach $\sim$130 K \citep{Spencer1999}, and visible and infrared spectra suggest that it is likely salty in composition \citep{Fischer2015, Fischer2017, TrumboEtAl2019}}. Yet, its 3.5 $\micron$ H$_2$O$_2$ band \edit1{area} reaches more than twice that of the most spectrally icy region. Figure \ref{fig:2016_spectra} shows spectra representative of both locations.

\begin{figure*}
\figurenum{2}
\plotone{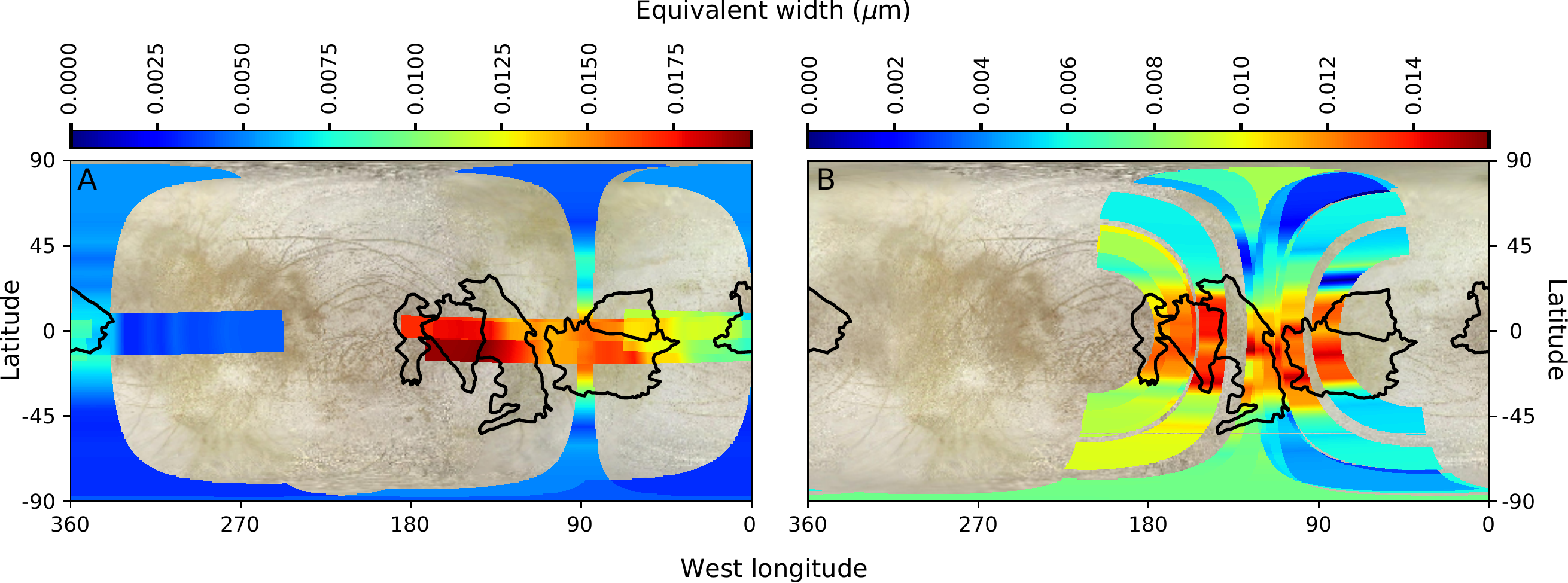}
\caption{All mapped slits from our 2016 (A) and 2018 (B) observations, which demonstrate the spatial distribution of Europa's 3.5 $\micron$ H$_2$O$_2$ absorption. The deepest absorptions map to low latitudes of the leading and anti-Jovian hemispheres and appear correlated with the geologically young chaos regions (outlined in black). The trailing/sub-Jovian slits, in comparison, display very weak absorptions, consistent with prior full-disk spectra \citep{HandBrown2013}.  \label{fig:maps}}
\end{figure*}

The apparent preference of H$_2$O$_2$ for low-latitude chaos terrain persists across all of the mapped 2016 slits (Figure \ref{fig:maps}A). Like the N/S slit of Figure \ref{fig:2016_spectra}, the leading-hemisphere E/W slit positions also show strong absorptions across Tara Regio. In addition, they demonstrate a marked difference between low-latitude chaos and low-latitude plains terrain, displaying enhanced absorption across the other large-scale chaos regions of the leading/anti-Jovian hemisphere (also mapped approximately from \citet{Doggett2009}), but much weaker band strengths across the ridged plains immediately east of Tara Regio ($\sim$30$\degree$ W). In fact, even the resolution elements that include the narrow space between Tara Regio ($\sim$85$\degree$ W) and eastern Powys Regio ($\sim$125$\degree$ W) appear to exhibit somewhat weakened absorptions on average. The trailing/sub-Jovian regions, by comparison, show only very weak 3.5 $\micron$ bands, consistent with the full-disk data of \citet{HandBrown2013}. However, what little H$_2$O$_2$ is present still seems correlated with chaos, as suggested by the slight enhancement near the outlined western-most portion of Annwn Regio ($\sim$0$\degree$ N, 360$\degree$ W). We note, however, that accurately quantifying such weak absorptions in this region is difficult, especially given the substantially different continuum shape of spectra from the low-latitudes of the trailing/sub-Jovian hemisphere.

While the 2016 map certainly suggests that Europa's H$_2$O$_2$ favors chaos terrain, it is also potentially consistent with a simple preference for low latitudes combined with the known hemispherical trend observed previously in full-disk spectra \citep{HandBrown2013}. In addition,  the E/W slits, though approximately constant in latitude, still each cover a range of surface temperatures reflecting the early morning through late afternoon, which may impart unknown diurnal effects onto the longitudinal patterns we observe. To account for these caveats, our 2018 map (Figure \ref{fig:maps}B) is composed of only N/S slit positions. This orientation more concretely differentiates between a preference for low latitudes and a specific correlation with the chaos regions, which are asymmetric about the equator. N/S slits also have the advantage of capturing the latitudes within a given slit position at similar times of the local day, thus minimizing diurnal complications to the observed H$_2$O$_2$ pattern within a slit. 

Our 2018 dataset confirms the trends suggested by the 2016 map---Europa's H$_2$O$_2$ prefers warm, low-latitude, ice-poor chaos terrain to the comparatively cold and ice-rich regions of the upper latitudes. The 2018 N/S slit positions expand our coverage of the large-scale chaos regions of the leading/anti-Jovian hemisphere, and all individual slit positions consistently show the largest 3.5 $\micron$ absorptions within the outlined chaos terrain. \edit1{In most cases, the observed band areas increase sharply near the largescale chaos boundaries, reaching values up to three times as large as those immediately above or below the chaos terrain.} In all cases, especially those of Tara Regio and eastern Powys Regio, this requires asymmetric displacement south of the equator, which is inconsistent with a simple preference for low latitudes. In fact, the complete 2018 map indicates that most of Europa's H$_2$O$_2$ falls between $\sim$40$\degree$ S and 20$\degree$ N, in a latitudinal pattern that possesses similar N/S asymmetries to those of the overall pattern of leading/anti-Jovian chaos terrain. This observed distribution requires dominating influences on Europa's H$_2$O$_2$ other than temperature or water ice abundance. 

\section{Discussion}\label{sec:discussion}
Outside of ice availability and surface temperature, perhaps the most obvious potential factor governing the geographic distribution of H$_2$O$_2$ on Europa is the local radiation environment. Though the entirety of Europa's surface is irradiated by energetic particles, the energy flux and particle composition is geographically nonuniform \citep{Paranicas2001, Paranicas2009, Dalton2013, Nordheim2018}. In fact, the observed H$_2$O$_2$ enhancement at low latitudes on the leading hemisphere bears some resemblance to the leading hemisphere ``lens" pattern of bombardment by very high-energy ($\gtrsim$ 20 MeV) electrons, which move against the co-rotation direction of Jupiter's magnetic field \citep{Paranicas2009, Nordheim2018}. This lens centers on the leading hemisphere apex, weakens toward the mid latitudes, and tapers toward the sub- and anti-Jovian points.   However, laboratory irradiation experiments have yielded more efficient H$_2$O$_2$ production and higher H$_2$O$_2$ concentrations using energetic protons and ions \citep{LoefflerEtAl2006} than they have using electrons \citep{HandCarlson2011}. This is due to the higher linear energy transfer (LET) of ions \citep{LoefflerEtAl2006, HandCarlson2011}, which reflects a higher density of resulting excitations and ionizations in the ice. Although H$_2$O$_2$ production by 20+ MeV electrons has never been directly studied in the lab, the LET of electrons at these energies is still more than an order of magnitude less than that of the majority of energetic protons and ions striking the surface \citep{HandCarlson2011}. In contrast to the leading hemisphere electrons, energetic protons and ions bombard Europa much more uniformly \citep{Paranicas2002, Paranicas2009} and potentially even favor the high latitudes \citep{Dalton2013}. Indeed, the combined bombardment patterns of energetic ions and $\lesssim$20 MeV electrons, which primarily impact the trailing hemisphere \citep{Paranicas2001}, result in a relative sheltering of the leading-hemisphere equator in terms of total flux \citep{Paranicas2009, Dalton2013}. Thus, while the leading hemisphere electron lens may help produce the H$_2$O$_2$ at low latitudes, it is unlikely to be the sole explanation. Indeed, because the lens is expected to be symmetric across the equator, it certainly cannot account for the apparent correlation of H$_2$O$_2$ with chaos terrain.

The geographic association of H$_2$O$_2$ with chaos terrain instead may imply that compositional differences across the surface are a dominating factor in controlling H$_2$O$_2$ abundance. \edit1{Chaos regions are geologically young and extensively disrupted areas of the surface interpreted to reflect recent interaction with subsurface material \citep[e.g][]{Collins2009, SchmidtEtAl2011}. In fact, }leading-hemisphere chaos regions have previously been shown to be compositionally distinct from their surroundings \citep{Fischer2015, Fischer2017, TrumboEtAl2019}, likely reflecting the contributions of endogenous material rich in chloride salts. To our knowledge, the effects of salty ice or frozen brines on the radiolytic formation of H$_2$O$_2$ have not been studied, leaving us with no reason to expect H$_2$O$_2$ enhancements due to salt. However, the addition of electron-accepting contaminants, such as O$_2$ and CO$_2$, to the ice has been shown to enhance H$_2$O$_2$ yields \citep{MooreHudson2000}. In these cases, the CO$_2$ and O$_2$ molecules slow the destruction of newly formed H$_2$O$_2$ by scavenging destructive electrons created during the continued irradiation of the ice. In fact, it has been noted that very limited NIMS data of portions of the leading hemisphere show a correlation between Europa's surface CO$_2$ and its H$_2$O$_2$ \citep{Carlson2001,Carlson2009}, although the spectra and corresponding maps were never published. Critically, these data cover portions of Tara Regio and Powys Regio, and demonstrate a clear enhancement of CO$_2$ in these chaos regions relative to the intervening plains (R. W. Carlson personal communication). Our NIRSPEC datasets map the same regions at the large scale. If we extrapolate the NIMS correlations, then the presence of CO$_2$ at low latitudes and in chaos terrain may help explain our observed H$_2$O$_2$ distribution. Perhaps more excitingly, if CO$_2$ is truly a dominant factor in controlling the H$_2$O$_2$ distribution on Europa, then the widespread coverage of our H$_2$O$_2$ data may hint at a broad association of Europa's CO$_2$ with its geologically young chaos terrain, and therefore with an interior carbon source. However, \edit2{as no comprehensive map of Europa's CO$_2$ was ever made, and as the corresponding spectra were never published, more complete investigations of the distribution and abundance of Europa's CO$_2$} are certainly needed to address this hypothesis. 

While we suggest that the observed H$_2$O$_2$ distribution may reflect that of Europa's CO$_2$, it is worth addressing some caveats to this hypothesis. One potential complication is the apparent lack of significant H$_2$O$_2$ on the trailing hemisphere, which contains extensive chaos terrain \citep{Doggett2009} and potential spectral evidence for CO$_2$ \citep{HansenMcCord2008}, although the spectra are poor quality. While water ice abundance does not appear to play a significant role in the latitudinal distribution of H$_2$O$_2$ on the leading and anti-Jovian hemispheres, it is possible that the more dramatic contrast in ice-fraction between the leading and trailing hemispheres accounts for their observed differences in H$_2$O$_2$ abundance. The widespread sulfur radiolysis experienced on the trailing hemisphere may also contribute to these differences. SO$_2$, a minor product of the radiolytic sulfur cycle on the trailing hemisphere \citep{Carlson2002}, reacts efficiently with H$_2$O$_2$ in water ice at temperatures relevant to Europa, rapidly producing sulfate \citep{LoefflerHudson2013}. Thus, while the composition of the leading hemisphere may enhance local H$_2$O$_2$ concentrations, the composition of the trailing hemisphere may act to deplete them. 

Lastly, we would like to acknowledge the possibility that the 3.5 $\micron$ H$_2$O$_2$ band \edit1{areas} may not directly reflect relative H$_2$O$_2$ abundances, because they may also include the effects of geographically varying grain sizes. With only a single absorption band, it is impossible to disentangle the two effects. However, laboratory data suggest that H$_2$O$_2$ on Europa likely exists as isolated trimers (H$_2$O$_2$ $\bullet$ 2H$_2$O) within the surface ice \citep{LoefflerBaragiola2005}. It has been suggested that the ice grain size on Europa increases with latitude \citep{Carlson2009, CassidyEtAl2013}, which should produce the opposite latitudinal trend in band \edit1{area} from what we observe. Thus, we feel that the 3.5 $\micron$ band of H$_2$O$_2$ likely represents a good proxy for its abundance.

\section{Conclusions}\label{sec:conclusions}
Using ground-based L-band spectroscopy, we have mapped Europa's 3.5 $\micron$ H$_2$O$_2$  absorption across its surface. Although laboratory data would seem to suggest that Europa's H$_2$O$_2$  should be most abundant at the cold, icy high latitudes of the leading hemisphere, we find the strongest absorptions at low latitudes, preferentially within geologically young chaos terrain. We suggest that this distribution may largely reflect efficient electron scavenging by CO$_2$ within the chaos regions, which can enhance H$_2$O$_2$ concentrations. This hypothesis would benefit from continued mapping of Europa's H$_2$O$_2$ and CO$_2$ and from further laboratory experiments involving ion and electron irradiation of ice mixtures potentially reflective of the composition of leading-hemisphere chaos terrain.  

\acknowledgements
This research was supported by Grant 1313461 from the National Science Foundation. This work was also supported by NASA Headquarters under the NASA Earth and Space Science Fellowship Program Grant 80NSSC17K0478. K. P. H. acknowledges support from the Jet Propulsion Laboratory, California Institute of Technology, under a contract with the National Aeronautics and Space Administration and funded in part through the internal Research and Technology Development program. The data presented herein were obtained at the W. M. Keck Observatory, which is operated as a scientific partnership among the California Institute of Technology, the University of California, and the National Aeronautics and Space Administration. The Observatory was made possible by the generous financial support of the W. M. Keck Foundation. The authors wish to recognize and acknowledge the very significant cultural role and reverence that the summit of Mauna Kea has always had within the indigenous Hawaiian community. We are most fortunate to have the opportunity to conduct observations from this mountain. We thank Robert W. Carlson for sharing enlightening details concerning \textit{Galileo} NIMS observations of CO$_2$ on Europa.

\software{Astropy \citep{Astropy}, skimage.transform \citep{skimage}}

\end{document}